\documentclass{elsart}

\usepackage{natbib}

\usepackage{graphics}
\usepackage{amssymb}

\newcommand{\kms}{km s$^{-1}$}

\begin{document}

\begin{frontmatter}



\title{Molecular gas in Arp 94: Implications for intergalactic star formation}


\author{Ute Lisenfeld}
\address{Dept. F\'\i sica Te\'orica y del Cosmos, Universidad Granada, Spain and
Instituto de Astrof\'\i sica de Andaluc\'\i a (IAA/CSIC), Apdo. 3004, 18080 Granada, Spain}

\author{Carole Mundell \& James Allsopp}

\address{Astrophysics Research Institute, Liverpool John Moores University, Twelve
 Quays House, Egerton Wharf, Birkenhead, CH41 1LD, U.K.}

\author{Eva Schinnerer}
\address{Max-Planck-Institut f\"ur Astronomie,
K\"onigstuhl 17, 69117 Heidelberg, Germany}

\begin{abstract}
We present $^{12}$CO(1-0) observations of the interacting galaxy system Arp 94,
which contains the Seyfert galaxies NGC~3227 and NGC~3226 as well
as the star-forming candidate dwarf galaxy J1023+1952.
We mapped the CO distribution in J1023+1952 with the IRAM 30m telescope
and found molecular gas across the entire extent of the neutral
hydrogen cloud -- an area of about 9 by 6 kpc.
The region where star formation (SF) takes place is restricted to 
a much smaller ($\sim$ 1.5 by 3 kpc) region in the south where the
narrow line width of the CO shows that the molecular gas
is dynamically cold. 
Neither the molecular nor the total gas surface density  in the
SF region are significantly higher than in the rest of the object suggesting that an external 
trigger is causing the SF.
The fact that CO is abundant and apparently a good tracer for
the molecular gas in J1023+1952 indicates that its metallicity is relatively high
and argues for a tidal origin of this object. 
%The large amount
%of molecular gas has most likely been stripped together
%with the atomic gas from relatively inner regions of NGC~3227
%during a second encounter between NGC~3227 and NGC~3226.
\end{abstract}

%\begin{keyword}


%\end{keyword}

\end{frontmatter}

% main text
\section{Some background on Arp 94}
\label{background}

Arp 94 is an interacting Seyfert system \citep[][2004]{mundell95},
consisting of the disturbed Seyfert galaxies NGC~3227, an SAB(s) pec barrred
spiral, and its elliptical companion NGC~3226 (E2 pec). 
Neutral hydrogen (HI) imaging of the system \citep{mundell95}
revealed two gaseous tidal tails extending 
 $\sim$ 100 kpc north and south
of NGC~3227, well-ordered gas  in the disk of NGC~3227 and a massive HI cloud
that lies apparently at the base of the northern tail and is close to, but
physically and kinematically distinct from, the disk of NGC~3227.
Mundell et al. (1995) suggested that this cloud (hereafter J1023+1952)
might be a dwarf galaxy that is either pre-existing and being accreted 
by Arp~94, or a 
newly-created Tidal Dwarf Galaxy (TDG) forming from the tidal debris.

The subsequent discovery of a region of very blue star-forming knots embedded
in a high HI column density ridge in the southern
 half of  J1023+1952  \citep{mundell04}
confirmed its classification as a dwarf galaxy. 
The inferred star formation rate (SFR), from
UV luminosities and H$\alpha$ equivalent widths, of 
the knots suggested a starburst 
age less than 10 Myr. Near infrared imaging provided 
further evidence for the youth
of J1023+1952, no additional embedded star formation (SF) or 
old stellar population were found
\citep{mundell04}, a result confirmed by recent infrared imaging with Spitzer
(Appleton et al., in prep.).

In this paper we present CO observations of  J1023+1952,
probing the distribution and kinematics of the molecular gas
in order to better understand the origin of this object and 
how SF proceeds in it. A more detailed description of the
observations, analysis and interpretation will be done in forthcoming
publication.

\section{CO observations and data analysis}
\label{data}

\begin{figure}
%\resizebox{0.5\hsize}{!}{\includegraphics[angle=270]{compare_ipac_gator_f12_lin_scale.ps}}
\resizebox{\textwidth]}{!}{\includegraphics{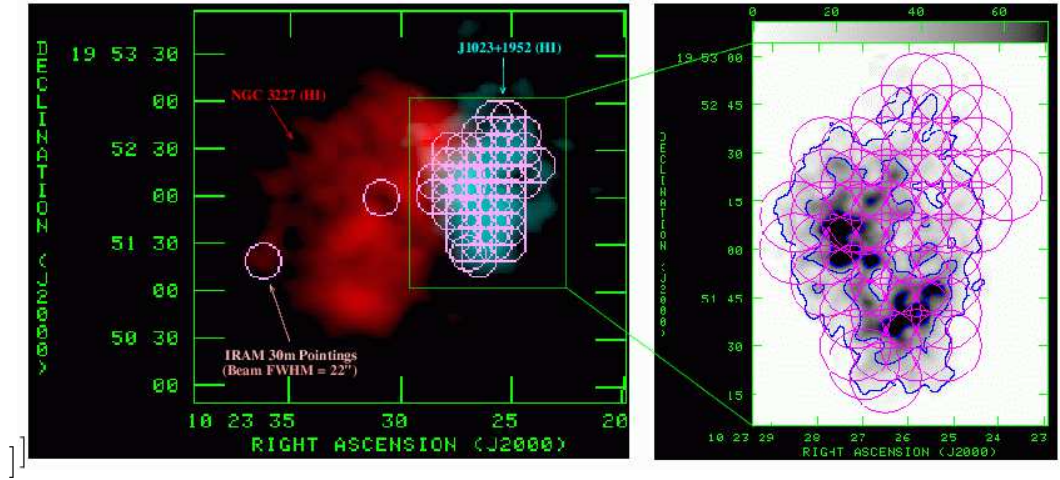}}
%\resizebox{0.5\hsize}{!}{\includegraphics{fig1a.ps}}
%\resizebox{0.5\hsize}{!}{\includegraphics{fig1b.ps}}
\caption{{\it Left:}
Location of IRAM 30m pointings (FWHM=22$^{\prime\prime}$) overlaid on colour image
of HI emission from spiral galaxy NGC~3227 (red) and  J1023+1952 (blue).
{\it Right:} Location of IRAM pointings on J1023+1952, overlaid
over HI emission.
}
\label{pointings}
\end{figure}

We observed the $^{12}$CO(1-0) and $^{12}$CO(2-1) (not shown here) lines in June 2004
with the IRAM 30-meter telescope at 115 and 230 GHz with a standard set-up.
We used the wobbler with a throw of 200$^{\prime\prime}$.
The location of the individual pointing, spaced 10$^{\prime\prime}$ apart within
J1023+1952, plus two additional pointings within NGC~3227,
are shown in Fig.~\ref{pointings}.
For the data reduction, we selected the observations taken 
during satisfactory weather
conditions, summed the spectra over the individual positions and subtracted a 
constant continuum level. 
At the edge of the galactic disk of NGC~3227 the emission of
this galaxy and J1023+1952
spatially overlaps in projection.  It has been possible to disentangle the different kinematic
components in the current $^{12}$CO(1-0) data cube by the fitting of a
double Gaussian line profile.

\section{Results}
\label{results}

\begin{figure}
\resizebox{1.2\textwidth}{!}{\includegraphics{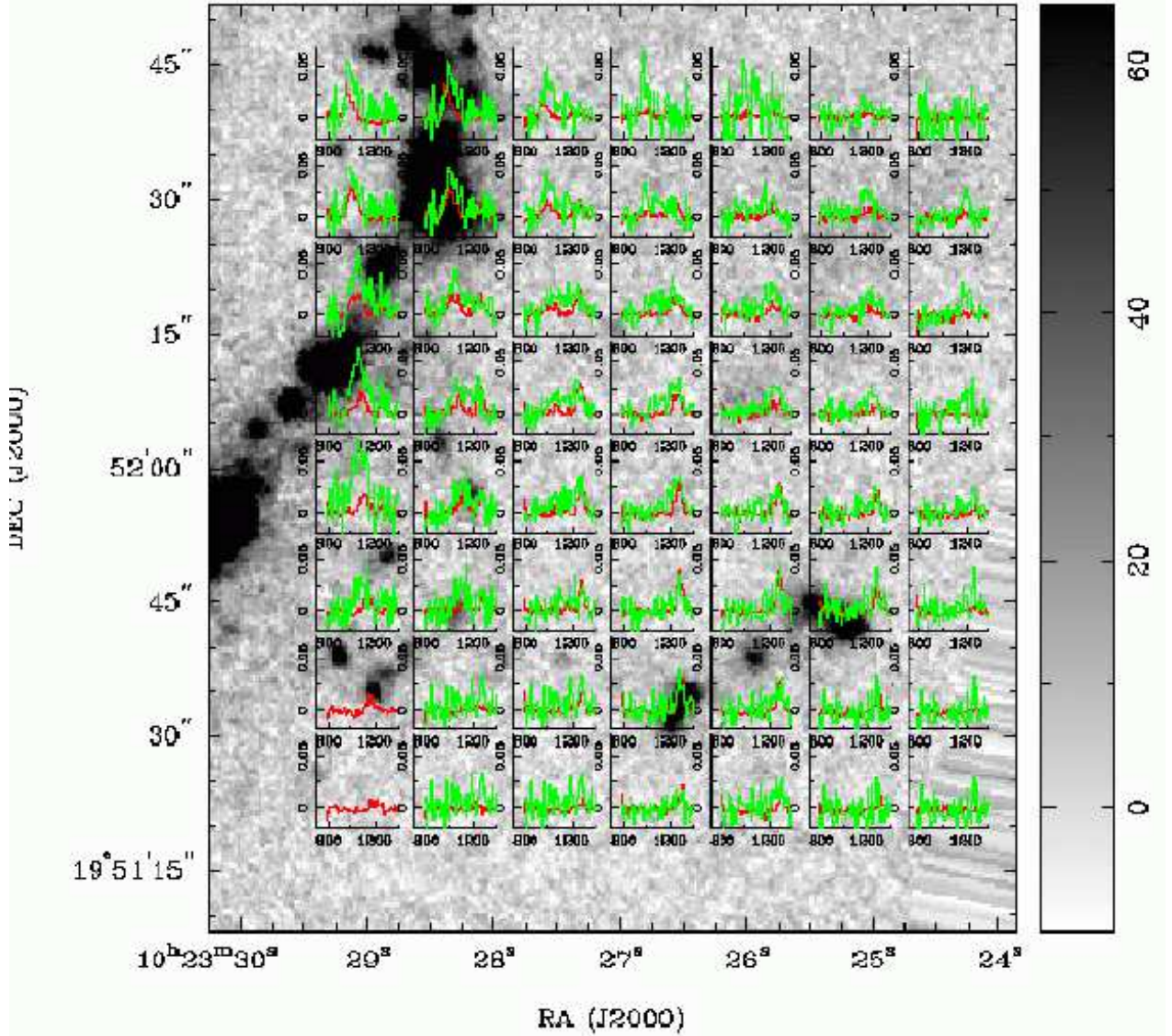}}
\caption{Spectra of HI emission (red) and CO(1-0) emission (green)
from  J1023+1952 overlaid on H$\alpha$ image from \citet{mundell04}.}
\label{spectra}
\end{figure}

\begin{figure}
\resizebox{0.32\textwidth}{!}{\includegraphics{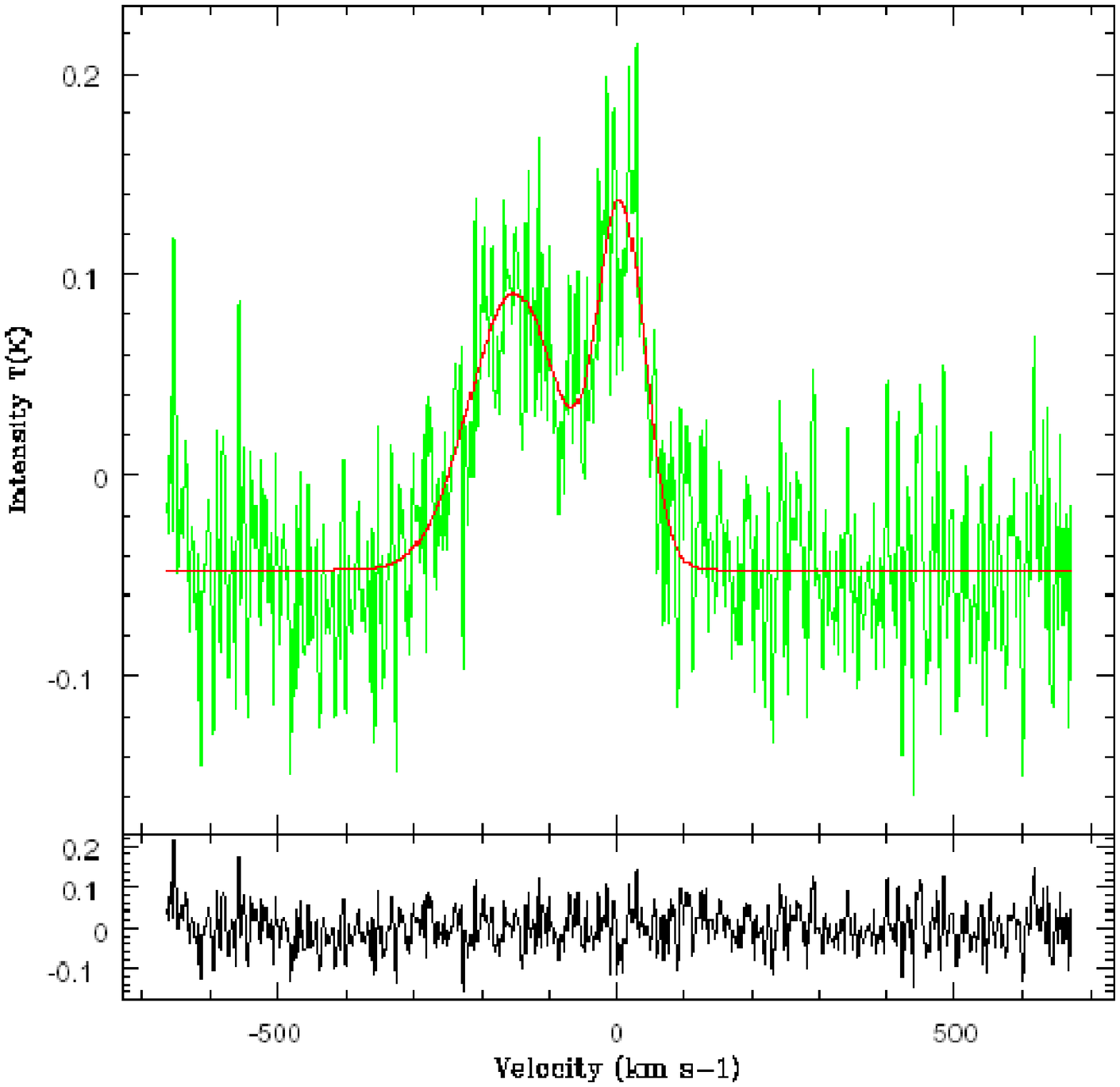}}
\resizebox{0.32\textwidth}{!}{\includegraphics{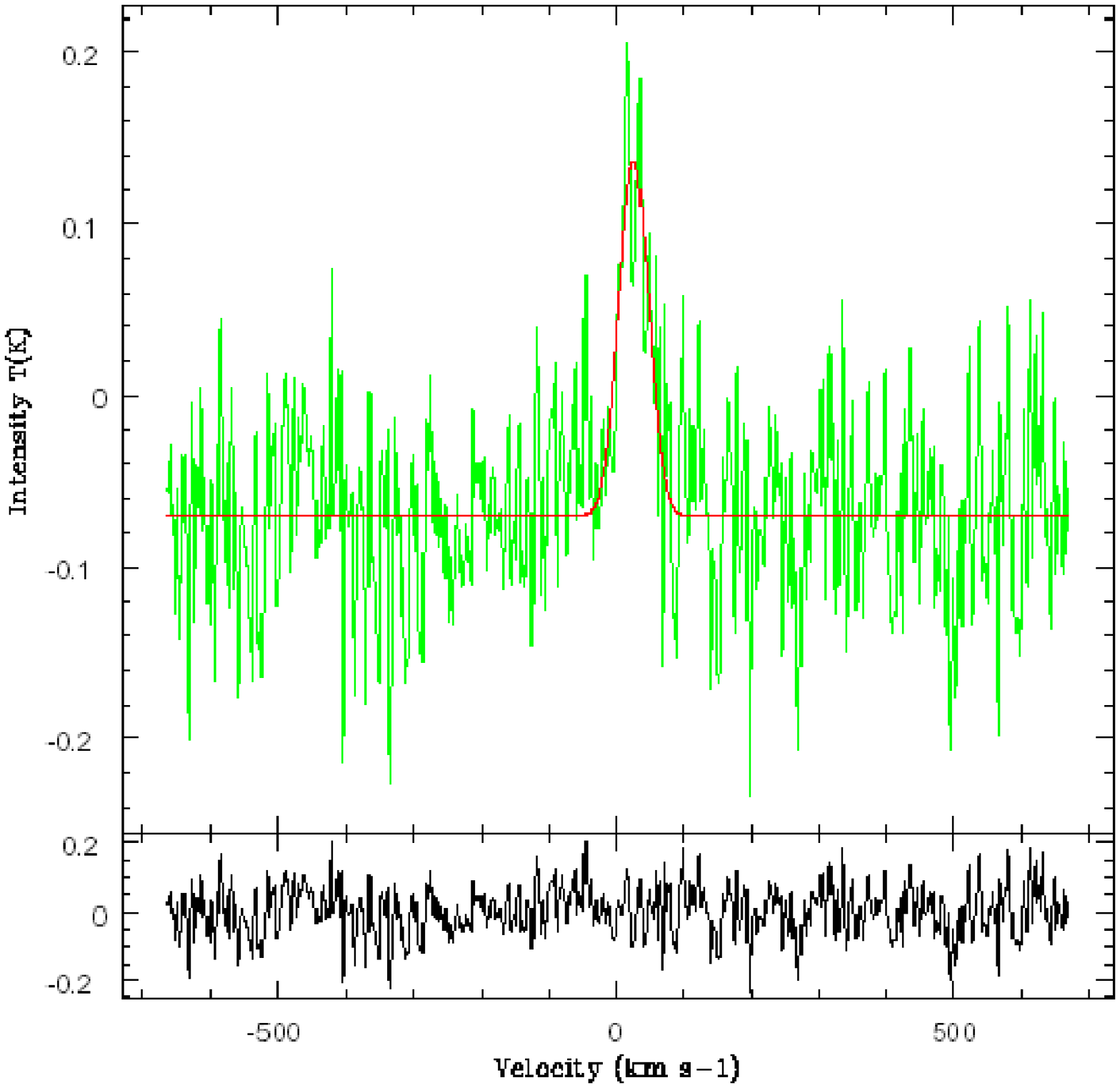}}
\resizebox{0.32\textwidth}{!}{\includegraphics{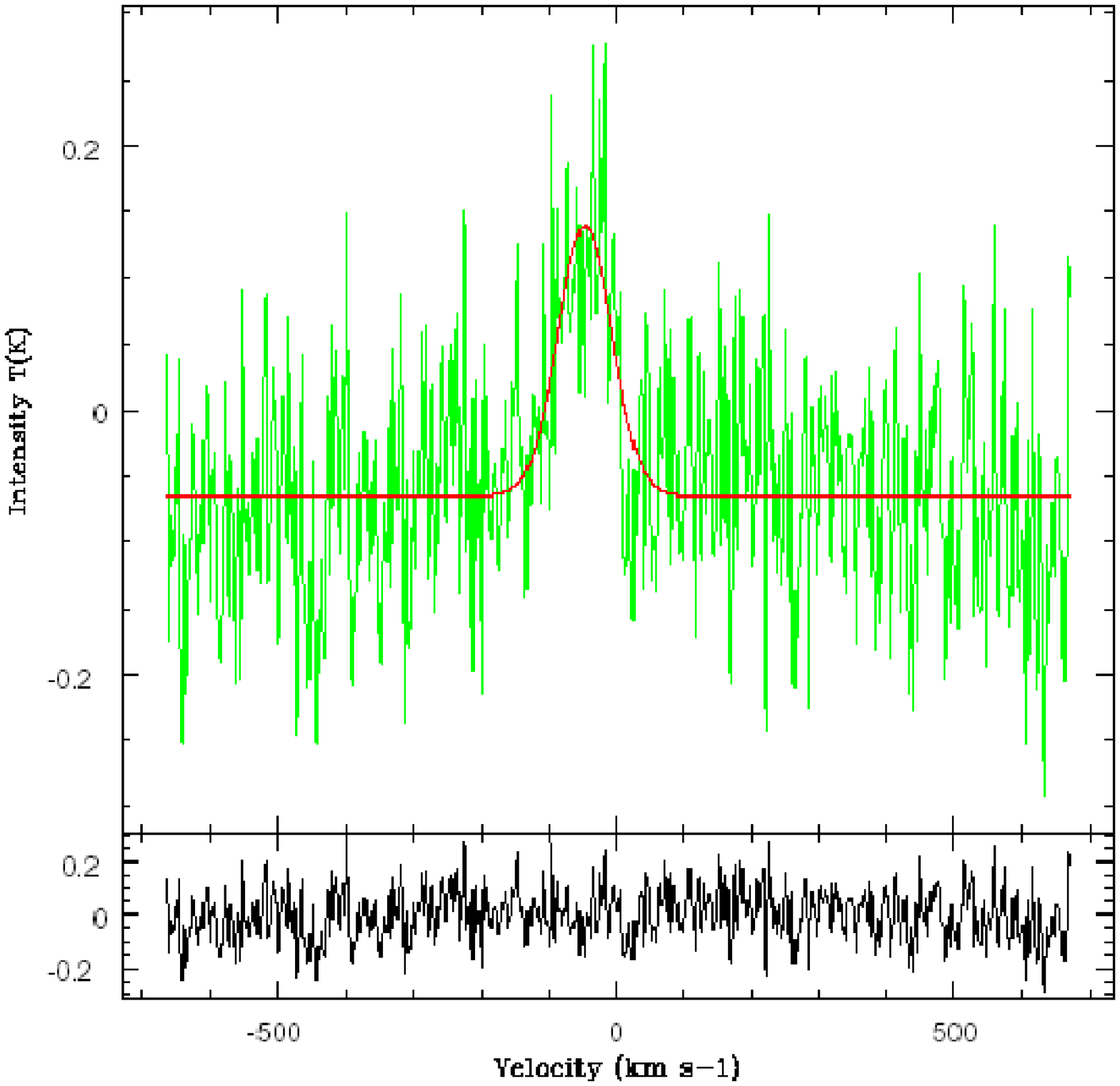}}
\caption{CO spectra together with gaussian fittings for 3 different regions:
{\it Left:} Region along the HI ridge close to NGC~3227 (around
position RA 10:23:28.0, DEC 19:52:00).
{\it Middle:} Region in the south with recent SF (around
position RA 10:23:25.5, DEC 19:51:40). 
{\it Right:} Region towards the north-west, with low HI emission (around
position RA 10:23:25.5, DEC 19:52:20).
}
\label{gauss_spectra}
\end{figure}

CO was detected at practically all positions.
In the nucleus and the outer disk of NGC~3227 strong CO emission was found.
In Fig.~\ref{spectra} we show the observed spectra in J1023+1952, overlaid with
HI spectra. The main results from the
observations can be summarized as follows: (i) CO was detected 
everywhere in the HI cloud, and not just in the southern part where
SF is taking place. The total extent of the
CO emitting region is $\sim$ 6 $\times$ 9 kpc. (ii) The line widths agree
well with those of HI. (iii) The CO line widths vary over the 
cloud: They are  narrow (FWZI of $30 - 40$
\kms) in the southern part 
coinciding with the SF region, and they are substantially broader in the
rest of the cloud (FWZI of $>$ 100 \kms).
The different line shapes are illustrated in Fig.~\ref{gauss_spectra}
where typical spectra from these 3 regions are shown together with the
corresponding Gaussian fits. 
%Apart from the different line widths, it
%can be seen that the spectra from NGC~3227 and J1023+1953 which spatially
%overlay towards the east, can be clearly separated. 
%
%Spectrum 1, obtained
%from a region along the HI ridge, close to NGC~3227, shows two components, 
%one corresponding
%to J1023+1953 and the other emitted from NGC~3227. The two components can
%be clearly separated with the Gaussian fit. Spectrum 2, obtained in the 
%north-western
%part of the cloud, in a region with a low HI content, shows a relatively broad
%line, whereas Spectrum 3, obtained from the SF region shows a much narrower 
%line.

%The total molecular mass derived from the CO, using a Galactic 
%conversion factor
%of $2 10^8$ K \kms\ cm$^{-2}$, is $2\times 10^8$ \msol, roughly half
%the HI mass ($5.1\times 10^8$ \msol, Mundell et al. 1995).

\section{Discussion}
\label{discussion}

\subsection{Triggering of the SF in the south}

The SF is restricted to a small region in the southern part. 
The absence of SF in the rest of the cloud is not due
to a lack of molecular gas because CO has been
found over the entire object.
Neither is the surface density of the molecular gas the 
parameter determining the distribution of SF,
because it
is quite uniform over the cloud, 
with values between 3 and 6$\times 10^{20}$ cm$^{-2}$.
Furthermore, the highest molecular gas surface
density does  {\it not} coincide with the SF region.
The surface density of the total (atomic and molecular) gas is
rather homogeneous as well with
values ranging between  1.5 and 4$\times 10^{21}$ cm$^{-2}$. 
Although the gas surface
density in the SF region is close to the highest values, 
other places along the HI ridge have similarly high values, so
that high gas surface density alone cannot be an explanation
for SF occuring in just one restricted area.

The only noticable difference in the gas properties is the
narrower line width in the region where SF takes place in comparison
to the rest of the cloud, indicating that the gas is 
dynamically cooler here.
The SF is triggered only in
the {\em dynamically-cold} molecular gas in the south of the cloud,
showing that overall gas-richness is not a sufficient condition for
extragalactic star formation. We suggest that the cold ridge and its
associated SF are likely externally triggered by the tidal
interaction.
%
%The fact that cold molecular gas and SF are only present in
%the south of the cloud, being the rest of it nevertheless gas 
%rich, argues for an external trigger for the SF, probably related
%to the interaction.

\subsection{Nature of J1023+1953}

It has been an open question whether  J1023+1953 is a preexisting dwarf galaxy,
involved in the interaction, or an object created out of tidal debris stripped
from the disk of NGC~3227, 
%which might evolve into a dwarf galaxy,
a candidate TDG.
%\citep[e.g.][]{mirabel92}.
One property that distinguishes dwarf galaxies from 
TDGs is their metallicity.
Since the latter are made from recycled gas, their metallicities
are close to those of the parent galaxies. Therefore, 
TDGs do not follow the magnitude-luminosity relation found for
other dwarf galaxies, and their metallicities lie
in a narrow range of 12+log(O/H) = 8.3-8.6 \citep{duc00}.

Unfortunately, the metallicity in J1023+1953 has not been measured yet.
However, the detection of large quantities of CO argues for a high
metallicity.
%well above the value expected from the blue magnitude of
%$M_{\rm B} = 13.5$ mag. 
The conservative estimate of J1023+1952's total brightness $M_B \sim
-15.5$, assuming an extinction of $A_V = 2$mag, (Mundell et al 2004)
 would predict a metallicity of  12+log(0/H) $\sim$ 8.
%The blue
%From the magnitude-metallicity relation of
%dwarf galaxies we expect a metallicity of 
%12+log(O/H) = 7.6-7.8 for J1023+1953. 
At these low metallicities the detection rate of dwarf in CO is
very low  \citep[e.g.][]{leroy05},
indicating that CO is no longer a good tracer of the molecular gas content.
Thus, the fact that we detect abundant molecular gas, as usually found
in TDGs   \citep{braine01},
%and that CO 
%seems to be a good tracer of the molecular gas, as found in
%TDGs \citep{braine01}, 
suggests that
the metallicity is close to the values found in spirals and
strongly argues for a tidal origin of J1023+1953.

%\subsection{Origin of molecular gas}

%The large amount of the molecular gas and the fact that its kinematics agrees
%very well with that of HI argues in favor of simultaneous stripping
%of both gas components from NGC~3227. In-situ formation of the molecular as
%an alternative process, suggested to take place in other TDGs
%(Braine 2000, 2001), is not likely in this object because of
%large extent of the molecular gas.

%Since molecular gas in spiral galaxies is restricted, in contrast to
%HI, to the inner parts, this implies that the gas (both molecular
%and atomic) has been stripped from relatively inner parts of NGC~3227.
%Since in a first encounter between NGC~3226 and NGC~3227 preferentially
%the outer, HI rich, parts have been stripped, this implies that
%both galaxies are experiencing at least their second encounter at the
%moment.

%\appendix

% \label{}

% Bibliographic references with the natbib package:
% Parenthetical: \citep{Bai92} produces (Bailyn 1992).
% Textual: \citet{Bai95} produces Bailyn et al. (1995).
% An affix and part of a reference:
%   \citep[e.g.][Ch. 2]{Bar76}
%   produces (e.g. Barnes et al. 1976, Ch. 2).

\end{document}